\documentstyle[aps]{revtex}
\begin{document}
\draft
\title{Dualist interpretation of quantum mechanics}
\author{G. Potvin}
\address{Defence Research Establishment Valcartier, Val-B\'{e}lair, Qu\'{e}bec, Canada, G3J 1X5}
\date{\today}
\maketitle
\begin{abstract}
An interpretation of quantum mechanics is proposed which augments the stochastic pilot-wave model, introduced by Nelson [E. Nelson, Phys. Rev., {\bf 150,} 1079 (1966)], with a dual guidance condition.
Namely, in addition to the stochastic guidance condition which describes how the wave controls the particle (wave-to-particle guidance condition), we introduce another stochastic guidance condition describing how the particle guides the wave (particle-to-wave guidance condition).
We therefore introduce an action-reaction principle in the pilot-wave formulation of quantum mechanics.
The particle-to-wave guidance condition takes the form of spontaneous transitions of the wavefunction that are Poisson distributed in time.
The wavefunction selected by the transition is influenced by its likelyhood with respect to the particle position, as well as certain conservation constraints.
It is shown that the stochastic particle-to-wave guidance condition reproduces certain aspects of the spontaneous localization model of Ghirardi, Rimini and Weber [G. C. Ghirardi, A. Rimini and T. Weber, Phys. Rev. D, {\bf 34,} 470 (1986)], and of Milburn's intrinsic decoherence model [G. J. Milburn, Phys. Rev. A, {\bf 44,} 5401 (1991)], while avoiding some of the difficulties associated with them.
The macroscopic limit of the dualist interpretation is analyzed for the case of a free body.
Possible improvements of this interpretation are discussed. 
\end{abstract}

\section{Introduction}

The main conceptual problem in quantum mechanics remains the link between the quantum and the classical worlds.
The wavefunction in quantum theory typically spreads out over a range of possible values of an observable property of a system, such as the position of a particle.
Furthermore, the Schr\"{o}dinger equation that describes the evolution of the wavefunctions is linear and deterministic, which implies that even macroscopic objects may not possess a determinate position, momentum, or any other property.
And yet, some way must be found to account for the impression we have that macroscopic objects have determinate positions and velocities at all times.
Much has been written over the years on possible solutions to this problem, particularly on a solution based on what is known as the orthodox or the Copenhagen interpretation \cite{Lande65,Despagnat76,Bell87,Bell90,Peierls91,Gottfried91,Cushing94,Whitaker96,HealeyHellman98,Goldstein98}.
The orthodox interpretation relies on the projection postulate, also known as the collapse of the wavefunction, to provide the required link between the wavefunction and what is actually measured in experiments.
The postulate states that given an initial wavefunction of a quantum system in a superposition of eigenstates of an observable property, it will collapse to one of those eigenstates whenever it undergoes a measurement by a macroscopic and classical instrument.
The wavefunction after the measurement is therefore the eigenstate corresponding to the eigenvalue of the observable which is obtained by the measurement.
The probability of obtaining a given eigenvalue is equal to the square modulus of the coefficient of the normalized initial wavefunction for that particular eigenvalue-eigenstate (see Ref. \cite{Despagnat76} for a complete account of the orthodox interpretation).

The questions that this account of quantum mechanics entails are well known.
The wavefunction collapse is a non-unitary, irreversible process, which, on the face of it at least, is completely at odds with the unitary, reversible evolution described by the Schr\"{o}dinger equation.
Also, the orthodox account of a quantum measurement made use of a dichotomy between microscopic quantum systems and macroscopic classical instruments.
When quantum systems evolve in isolation, they do so according to Schr\"{o}dinder's equation, but when they interact with a classical instrument, their wavefunction collapses.
And yet, are not the macroscopic instruments made up of microscopic quantum particles?
If the Schr\"{o}dinger equation is truly universal, this would mean that the instruments pointer should be in a superposition of macroscopically distinguishible states after the measurement interaction.
What would such a result mean?
Alternatively, if the Schr\"{o}dinger equation is not universal and collapses do occur, there still remains the question of under what conditions they occur.
It is the resolution of this `measurement problem' which is the focus of the different interpretations of quantum mechanics.

The dualist interpretation attempts to solve the measurement problem, while avoiding the difficulties of previous attempts, by augmenting a pilot-wave interpretation with an action-reaction principle.
We will therefore begin with a review of some pilot-wave interpretations, along with a few other interpretations, in Sec. \ref{S_Rev}.
The review is by no means exhaustive.
We will only examine those relevant to the dualist interpretation, which will be explained in Sec. \ref{S_DI}.
Discussions on the relative merits of the dualist interpretation with respect to the others, along with areas still to be explored, are given in Sec. \ref{S_Disc}.
We conclude in Sec. \ref{S_Concl}.

\section{Review of some Interpretations of Quantum Mechanics}
\label{S_Rev}

As mentioned before, this review does not include all attempts to answer the measurement problem.
Specifically, we will not look at what Bub \cite{Bub97} called the `new orthodoxy', which is an interpretation built on notions of consistent histories, quantum logic and environment-induced decoherence and claims to do away with the need for a wavefunction collapse to describe the classical world (see Omn\`{e}s \cite{Omnes92}, Zurek and Paz \cite{ZurekPaz95}, Zurek \cite{Zurek91} and the corresponding comments and reply \cite{Anderson.et.al93}).
Also omitted are the statistical \cite{Ballentine.et.al70} and relative state interpretations \cite{Everett57} of quantum theory, among others (see Ref. \cite{Whitaker96} for a review).

\subsection{Pilot-wave interpretations}
\label{Sb_PW}
\subsubsection{Deterministic version}
\label{Ssb_Dv}
Originally proposed by de Broglie \cite{Cushing94} and later revived by Bohm \cite{Bohm52}, the pilot-wave approach states that in addition to the wavefunction propagating through space, there also exists a particle guided by the wave and moving along a well defined trajectory.
For a single nonrelativistic, spinless particle of mass $m$, we have
\begin{equation}
i \hbar \frac{\partial \psi}{\partial t} = -\frac{\hbar^{2}}{2m} \nabla^{2} \psi + V \psi
\label{E_Sch}
\end{equation}
\begin{equation}
\frac{{\rm d} {\bf x}_{p}}{{\rm d} t} = \frac{\hbar}{2mi} \frac{\psi^{*} {\bf \nabla} \psi - \psi {\bf \nabla} \psi^{*}}{| \psi |^{2}} \bigg{|}_{{\bf x}={\bf x}_{p}}
\label{E_vp}
\end{equation}
where $\psi = \psi({\bf x},t)$ is the wavefunction evolving under the influence of a potential $V$, and ${\bf x}_{p} = {\bf x}_{p}(t)$ is the trajectory of the particle.
Both the wavefunction and the particle are considered real, even though the wavefunction exists in configuration space for a many-body system.
We will therefore say that the pilot-wave interpretation possesses a {\it dual ontology}.
A {\it complete} description of an individual quantum system must therefore include an initial wavefunction and an initial particle postition.
It is convenient to express the wavefunction in polar coordinates, $\psi = R \ {\rm exp}[iS/\hbar]$, and to rewrite Eqs. (\ref{E_Sch}) and (\ref{E_vp}).
\begin{equation}
\frac{\partial S}{\partial t} + \frac{({\bf \nabla} S)^{2}}{2m} + V - \frac{\hbar^{2}}{2m}\frac{\nabla^{2}R}{R} = 0
\label{E_HJQ}
\end{equation}
\begin{equation}
\frac{\partial P}{\partial t} + {\bf \nabla}\cdot(P {\bf v}) = 0
\label{E_Cont}
\end{equation}
\begin{equation}
\frac{{\rm d}{\bf x}_{p}}{{\rm d} t} = \frac{{\bf \nabla}S}{m} \bigg{|}_{{\bf x}={\bf x}_{p}}
\label{E_vp2}
\end{equation}
where $P = R^{2}$ and ${\bf v} = {\bf \nabla}S/m$.
Equation (\ref{E_HJQ}) is a classical Hamilton-Jacobi equation with an additional term depending on the form of the $R$-field (called the `quantum potential' by Bohm \cite{Bohm52}, which imposes a corresponding `quantum force' on the particle).
Equation (\ref{E_Cont}) is a conservation equation for $P$ with a velocity field that also serves to guide to particle in Eq. (\ref{E_vp2}).

We will call the above the {\it Deterministic Pilot-Wave} (DPW) interpretation, although it has also been referred to in the literature as the de-Broglie-Bohm interpretation, the causal interpretation, the ontological interpretation, or even as Bohmian mechanics.
Recently, there has been a resurgence of interest in the DPW interpretation, resulting in a series of articles reformulating well known quantum mechanical problems, such as Heisenberg's uncertainty principle \cite{Robinson80}, the double-slit experiment \cite{Home.et.al89}, quantum fields and gravitation \cite{Holland.et.al88}, the correspondence principle \cite{Bohm.et.al85}, spin measurements and the EPR experiment \cite{Dewdney.et.al87}, the harmonic oscillator \cite{Dewdney.et.al84}, relativistic quantum mechanics \cite{Kyprianidis85,Berndl.et.al96}, and culminating in three major works on the subject \cite{BohmHiley93,Holland93,Cushing.et.al96}(see Ref. \cite{Berndl.et.al95} for a concise overview).
The DPW interpretation also allows us to speak clearly about arrival time distributions \cite{LeavensFinkelstein93}, something which the orthodox interpretation has difficulty defining.
The main advantage of DPW is its answer to the measurement problem: the particle under study and the instrument's pointer have determinate positions at all times, the measurement interaction induces a correlation in the combined particle-pointer wavefunction, which, by the guidance condition (Eq. (\ref{E_vp2})), also induces a correlation between the particle and pointer positions.
Contrary to the orthodox interpretation, the wavefunction never undergoes a collapse.
Equations (\ref{E_Sch}) and (\ref{E_vp}) entirely describe the dynamics of the wave-particle system.
DPW is therefore a no-collapse interpretation, as well as possessing nonlocal `hidden variables'.
The hidden variables are the particle positions, and they are nonlocal because a change in one particle position can instantly induce a change in another distant particle entangled with the first.

It is immediately apparent, however, that the probability density of the particle position, $\rho({\bf x}_{p})$, is logically distinct from the square modulus of the wavefunction, $P$.
The statistics of quantum theory are only reproduced in DWP theory if $\rho = P$, a condition D\"{u}rr {\it et al.} \cite{Berndl.et.al96,Durr.et.al92} called `quantum equilibrium' (better known as Born's statistical postulate).
The fact that the particle velocity is the same as the velocity in the $P$-field conservation law (Eq. (\ref{E_Cont})), ${\rm d}{{\bf x}}_{p}/{\rm d}t = {\bf v}({\bf x}_{p})$, means that if quantum equilibrium holds for a given time, it will hold for all times.
Of course, the opposite is also true, as was pointed out by Keller \cite{Keller53}.
One answer to this problem is to simply postulate quantum equilibrium.
Alternatively, one can invoke the fact that for a complicated system with a large number of degrees of freedom, the motion of the particles will be sufficiently complicated as to produce an effective mixing and diffusion of $\rho$, in a coarse-grained sense \cite{Cushing94,BohmHiley93,Bohm53,Valentini91}.
An initial probability distribution not in quantum equilibrium will eventually reach it, resulting in what Valentini called a subquantum H-theorem \cite{Valentini91}.
The use of probability in DPW theory is therefore no better or worse than the use of probability in classical statistical mechanics (see Sklar \cite{Sklar93} for a lucid account of the conceptual difficulties in classical statistical mechanics).
It is interesting to note that the guidance condition, Eq. (\ref{E_vp2}), does not uniquely conserve quantum equilibrium.
As Deotto and Ghirardi \cite{DeottoGhirardi98} showed, it is possible to add an additional velocity field to the guidance condition that still preserves quantum equilibrium.
We therefore have many possible guidance conditions in the DPW interpretation which can reproduce the predictions of orthodox quantum theory, although the one described by Eq. (\ref{E_vp2}) is the simplest.

Although the DPW interpretation includes particle trajectories, similar to the trajectories in classical physics, it is by no means guaranteed that very large and massive bodies would exhibit classical trajectories \cite{Holland93,Appleby99a}.
In other words, even though the quantum potential scales as $m^{-1}$, it still may show a singular behaviour in regions where $R \rightarrow 0$, causing significant deviations from the expected classical behaviour.
According to Bohm and Hiley \cite{BohmHiley93}, and Appleby \cite{Appleby99b}, the interaction with a random environment induces random fluctuations in the quantum potential such that classical trajectories are recovered to a good approximation.
Since the evolution of the wavefunction is rigorously deterministic, however, the randomness of the environment cannot be derived from the dynamics, but must be assumed (i.e. as reflecting our ignorance of its many degrees-of-freedom).

Furthermore, should the wavefunction be split into two or more non-overlapping wave packets in a measurement process, the particle will be seen to be in one of the packets.
But since the wavefunction is considered objectively real, the empty wave packets will continue to exist after the measurement, a situation deemed awkward by Stapp \cite{Stapp93}.
In the orthodox interpretation, the measurement only allows the packet corresponding to the eigenvalue which is obtained in the measurement, all other packets are eliminated.
In the DPW interpretation, the empty packets are not eliminated and might interfere with the occupied packet at a later time.
Bohm and Hiley \cite{BohmHiley93,BohmHiley85} invoke the interaction of the system with the measurement apparatus, and the interaction of the measurement apparatus with the environment, to argue that the empty packets rapidly become incapable of effectively affecting the occupied packet, thereby reproducing the measurement process in the orthodox interpretation.
Other objections to the DPW interpretation refer to its lack of symmetry, i.e. position and momentum are no longer on the same footing, position now being a preferred variable \cite{Lande65}.
Also, the wave acts on the particle but not the other way around \cite{Holland93}.
As we will see, the last objection does not apply to the dualist interpretation.

\subsubsection{Stochastic version}
\label{Ssb_Sv}
A variant on the DPW interpretation is what we will call the {\it Stochastic Pilot-Wave} (SPW) interpretation.
Here, as in the DPW theory, the particle and the wavefunction are both objectively real, and the evolution of the wavefunction is described solely by the deterministic Schr\"{o}dinger equation.
The motion of the particle, however, is described by a stochastic process which is conditioned by the wavefuncton.
Inspired partly on the idea by Bohm and Vigier \cite{BohmVigier54} of a subquantum fluid imparting random fluctuations on the particle, Nelson \cite{Nelson66} formulated a SPW theory in terms of a continuous Markov process.
The stochastic process of the particle is described by the Langevin equation (see Gillespie \cite{Gillespie96} for a concise account of Markov processes),
\begin{equation}
{\bf x}_{p}(t + {\rm d}t) = {\bf x}_{p}(t) + {\bf b}({\bf x}_{p},t){\rm d}t + D^{1/2} {\bf w}(t)({\rm d}t)^{1/2}
\label{E_Lgvn}
\end{equation}
where $D = \hbar / m$ is the diffusion constant, and ${\bf b}({\bf x}_{p},t)$ is the drift function given by
\begin{equation}
{\bf b} = \frac{{\bf \nabla}S}{m} + \frac{\hbar}{m} \frac{{\bf \nabla}R}{R} {\rm .}
\label{E_Drift}
\end{equation}
The first term is the particle velocity in Eq. (\ref{E_vp2}), while the second term is an osmotic velocity.
The function ${\bf w}(t)$ is an uncorrelated random time series with zero mean and unit variance:
\begin{mathletters}
\label{E_Nt}
\begin{equation}
<w_{i}(t)> = 0
\label{E_Nta}
\end{equation}
\begin{equation}
<w_{i}(t)w_{j}(t)> = \delta_{ij}
\label{E_Ntb}
\end{equation}
\begin{equation}
<w_{i}(s)w_{j}(t)> = 0, \qquad s \neq t
\label{E_Ntc}
\end{equation}
\begin{equation}
<x_{p,i}(s) w_{j}(t)> = 0, \qquad s \leq t 
\label{E_Ntd}
\end{equation}
\end{mathletters}
where the angle brackets denote an ensemble average.
Defining $\rho({\bf x}_{p},t|{\bf x}_{p0},t_{0})$ as the conditional probability density function for the particle position ${\bf x}_{p}$ at the time $t$, given that the position was ${\bf x}_{p0}$ at the earlier time $t_{0} < t$, it can be shown that Eqs. (\ref{E_Lgvn}) to (\ref{E_Nt}) result in a (forward) Fokker-Planck equation, 
\begin{equation}
\frac{\partial}{\partial t} \rho({\bf x}_{p},t|{\bf x}_{p0},t_{0}) + {\bf \nabla} \cdot [\,\rho({\bf x}_{p},t|{\bf x}_{p0},t_{0})\, {\bf b}({\bf x}_{p},t)] = \frac{D}{2} \nabla^{2} \rho({\bf x}_{p},t|{\bf x}_{p0},t_{0})
\label{E_FFPC}
\end{equation}
where the time and space derivatives are with respect to $t$ and ${\bf x}_{p}$.
Equation (\ref{E_FFPC}) would diffuse the conditional probability density function in such a way so that $\rho({\bf x}_{p},t_{0}+\tau | {\bf x}_{p0},t_{0}) \rightarrow P({\bf x}_{p},t_{0}+ \tau )$ as $\tau \rightarrow \infty$ \cite{BohmHiley93,BohmHiley89}.
Also, once quantum equilibrium is achieved, the osmotic velocity and diffusion terms cancel each other out.
Equation (\ref{E_FFPC}) then becomes identical to Eq. (\ref{E_Cont}), and so quantum equilibrium is conserved.
Given an initial probability density, $\rho({\bf x}_{p0},t_{0})$, the probability density at a later time, $\rho({\bf x}_{p},t) = \int \rho({\bf x}_{p},t | {\bf x}_{p0},t_{0}) \rho({\bf x}_{p0},t_{0}) {\rm d}^{3} x_{p0}$, also obeys a Fokker-Planck equation,
\begin{equation}
\frac{\partial}{\partial t} \rho({\bf x}_{p},t) + {\bf \nabla} \cdot [\,\rho({\bf x}_{p},t)\, {\bf b}({\bf x}_{p},t)] = \frac{D}{2} \nabla^{2} \rho({\bf x}_{p},t) {\rm .}
\label{E_FFP}
\end{equation}
Quantum equilibrium can therefore be seen as a natural consequence of the SPW interpretation, without the need for mixing or coarse-graining.
As with the DPW interpretation, it is possible to modify the stochastic guidance condition, Eq. (\ref{E_Lgvn}), in such a way as to attain and preserve quantum equilibrium \cite{Bacciagaluppi98}.
And although, under certain conditions, the most probable particle path is approximately classical \cite{Kirillov94}, there are also conditions where the SPW interpretation may exhibit significantly non-classical paths, even for very massive bodies.

Like its deterministic counterpart, the SPW interpretation has been the subject of numerous articles dealing with the physical meaning of the stochastic process \cite{Guerra.et.al83}, the extension of the theory to include spin \cite{Cohendet.et.al88}, quantum fields \cite{Lim.et.al85}, relativistic quantum mechanics \cite{BohmHiley89,Namsrai.et.al81} and mixed states \cite{Jaekel.et.al84}.
Some have taken the stochastic process as fundamental, and the wavefunction as a derived quantity.
However, as Wallstrom \cite{Wallstrom94} pointed out, the wavefunction must satisfy certain quantization conditions which cannot be derived from the components of the drift function alone.
Also, the statistics of a mixed state can only be reproduced if the stochastic process of each wavefunction in the ensemble is treated separately, thereby conferring a reality to the wavefunction \cite{Jaekel.et.al84}.
It is for these reasons that we classify stochastic process interpretations as pilot-wave interpretations.

\subsection{Explicit wavefunction collapse interpretations}
\label{Sb_WC}
An alternative to the introduction of a particle into the standard formulation of quantum mechanics is to add a mechanism, stochastic or deterministic, which causes the wavefunction to collapse.
The collapse mechanism is intended as an answer to the ill-defined measurement process of the orthodox interpretation by exactly specifying the conditions under which the collapse occurs.
Here, the wavefunction is assumed to completely describe an individual system.
There are no particles.
The wavefunction collapse mechanism, however, is not derived from the Schr\"{o}dinger equation, but is added to it.
The Explicit Wavefunction Collapse (EWC) interpretation, therefore, has a {\it monist ontology}, but {\it dual dynamics}.
Strictly speaking, however, the EWC is not an interpretation of quantum mechanics, but rather a rival theory (as is the dualist interpretation).
Nevertheless, we will still call the EWC an interpretation.

An influential model of EWC was introduced by Ghirardi, Rimini and Weber (GRW) \cite{Ghirardi.et.al86,Benatti.et.al87}.
It is this model which will be of greatest interest to us, mainly because of its simplicity.
GRW propose that the wavefunction undergoes spontaneous localizations events at random times.
Pearle called these events `hits' \cite{Pearle89}.
If the wavefunction of a single particle prior to a hit is $\psi({\bf x},t)$, the wavefunction undergoes the instantaneous collapse,
\begin{equation}
\psi({\bf x},t) \rightarrow (\alpha/\pi)^{3/4} F^{-1/2} \exp[-(\alpha /2)({\bf x}-{\bf z} \,)^{2}] \psi({\bf x},t)
\label{E_Hit}
\end{equation}
where the factor $F$ ensures that the altered wavefunction is correctly normalized, $\alpha^{-1/2}$ is the localization width, and ${\bf z}$ is the center of the hit.
The center of the hits are random but not equally likely: $F = F({\bf z} \,)$ is also equal to the probability density that the hit is centered at ${\bf z}$.
\begin{equation}
F({\bf z} \,) = \int (\alpha/\pi)^{3/2} \exp[-\alpha ({\bf x} - {\bf z} \,)^{2}] |\psi({\bf x},t)|^{2} {\rm d}^{3} x
\label{E_Nz}
\end{equation}
It is clear from Eq. (\ref{E_Nz}) that $\int F({\bf z} \,) {\rm d}^{3} z = 1$.
The hits are Poisson distributed in time with a frequency parameter $\lambda$.
GRW set the localization width to $\alpha^{-1/2} \approx 10^{-5} {\rm cm}$, and the frequency to $\lambda \approx 10^{-8} {\rm yr}^{-1}$.
If we have a macroscopic object with $N \approx 10^{23}$ particles, GRW showed that the center-of-mass coordinate collapses whenever a single particle in the object undergoes a hit.
Therefore, the center-of-mass collapses with a frequency $N \lambda \approx 10^{7} {\rm s}^{-1}$.

This means that for a single particle, the hits occur so infrequently that one is unlikely to observe it in an experiment.
However, for a macroscopic object, the hits occur so frequently that it becomes very difficult to observe a coherent superposition of macroscopic states separated by a distance greater than the localization width.
Other models have been developed that use a Continuous Stochastic Localization (CSL) process, rather than the discontinuous hits of the GRW model, for particle position \cite{Pearle89}, mass density, c-numbers \cite{Diosi.et.al89}, and with relativistic features \cite{Pearle99}.
These models were criticized by Ballentine \cite{Ballentine91}, who emphasized that the energy production they entail is incompatible with equilibrium and steady states.
Furthermore, the GRW model has two new physical constants to evaluate, something which can be problematic \cite{SquiresPearle94}.
Indeed, since we have a monist ontology, the wavefunction must entirely account for the definitiveness of the macroscopic world.
Therefore, the hit frequency must be low enough so as to leave microscopic dynamics essentially unchanged from the usual Schr\"{o}dinger equation, but high enough so as to ensure that \emph{all} of what we might consider macroscopic variables (the position of a spec of dust, say) collapse sufficiently quickly before anybody can `notice'.
To the best of the author's knowledge, the energy production predicted by GRW has not actually been detected so far.
Only ambiguous or negative results were obtained, imposing limits or constraints on the possible values for the frequency parameter and localization width.
Also, the GRW model (like the pilot-wave interpretation) requires that all quantum measurements are essentially position measurements.
It is not clear that this should always be the case \cite{Wan.et.al91}, but we will assume it to be true for the purpose of this work.

The monist ontology is also the cause of what is known as the `tail' problem \cite{Ghirardi.et.al95} (also, see Ghirardi and Grassi in Ref. \cite{Cushing.et.al96}).
In orthodox quantum theory, a system can only be said to possess a determinate value, $a$, of an observable, $A$, if the c-number of the corresponding eigenfunction satisfies \emph{exactly} $|c_{a}|^2 = 1$, and all other c-numbers are \emph{exactly} zero.
If $|c_{a}|^2 = 1 - \epsilon$, where $0 < \epsilon \ll 1$, then the system is said to be indeterminate with respect to the observable $A$, \emph{no matter how small} $\epsilon$ \emph{may be}.
In the orthodox theory, this requirement for determinateness is satisfied automatically by the reduction postulate.
However, it is never satisfied in the GRW theory.
To see this, we will suppose that the initial wavefunction in Eq. (\ref{E_Hit}) is spread out over a macroscopically distinguishable distance $D \gg \alpha^{-1/2}$.
Clearly, then, the system does not initially possess a determinate position.
But the position does not become determinate as a result of the hit described by Eq. (\ref{E_Hit}).
This is because that, while the Gaussian responsible for the localization becomes very small far away from the center ${\bf z}$, it is never actually zero (i.e. it has infinitely long `tails').
It is possible to reformulate GRW in terms of localization functions that are not Gaussian \cite{Weber91}.
The tail problem would be partially remedied if we choose a localization function with a finite support.
But this tactic can only work for the discrete localization model, which does not preserve the symmetry of the wavefunction.

The CSL model does preserve the symmetry of the wavefunction, however.
The continuous stochastic process in CSL causes the wavefunction to exponentially approach a determinate state (i.e.: $|c_{a}|^{2} = 1 - \exp[-\gamma t]$).
But that determinate state is never actually attained in a finite time, and so the system will never possess a definite value of $A$.
Proponents of the EWC interpretation attempt overcome this difficulty by defining an `objective reality' and a `projective reality'.
The value $a$ of a system is said to be objectively real if $|c_{a}|^{2} > 1 - \epsilon_{0}$, where $0 < \epsilon_{0} \ll 1$ is an objective reality threshold and constitutes an additional parameter in the theory.
If this criterion is not satisfied, then a projection operator is applied, creating a field in space-time, $|c_{a}|^{2} = |c_{a}|^{2}({\bf x},t)$, that is considered projectively real.

\subsection{Intrinsic decoherence}
Another interesting modification to the Schr\"{o}dinger equation is Milburn's intrinsic decoherence model \cite{Milburn91}.
Without going into detail, the intrinsic decoherence model postulates that over sufficiently small time scales, a system evolves by a random sequence of unitary phase changes generated by the Hamiltonian.
These random phase changes consequently induce decoherence in the energy basis (note, however, that the wavefunction does not actually collapse).
See Ref. \cite{Moya-Cessa.et.al93} for some applications of intrinsic decoherence. 
This approach is `softer' that the GRW model since the constants of the motion remain constant (no energy production).
Similarly, Steane \cite{Steane90} proposed that the measurement problem is solved whenever a measurement-like process causes the phase of a quantum system to become formally undecidable.
Indeed, many authors believe \cite{Peierls91,Gottfried91,Omnes92,ZurekPaz95,Zurek91} that the measurement problem is solved in large part, if not completely solved, by decoherence in one basis or another.
However, according to Bell \cite{Bell90}, Holland \cite{Anderson.et.al93}, and Bub \cite{Bub97} among others (see Leggett, Healey and Elby in Ref. \cite{HealeyHellman98}), decoherence, whether intrinsic or induced by the environment, necessarily requires the introduction of additional (and usually tacit) assumptions about the meaning of the wavefunction to fully account for the definitiveness of the macroscopic world.
We now have enough of a background to formulate the dualist interpretation.

\section{The dualist interpretation}
\label{S_DI}

The idea for an action-reaction principle for the DPW interpretation was suggested by Holland \cite{Holland93}, but not pursued.
A more serious attempt was subsequently made by Abolhasani and Golshani \cite{AbolhasaniGolshani98}.
However, in their model, the wavefunction is directly coupled to the probability density of the particle position, and only indirectly on the particle position itself.
It is unclear, therefore, how the particle affects the wavefunction.
Also, Santos and Escobar \cite{SantosEscobar98} proposed a combination of a variant of the SPW interpretation (a `beable' interpretation) with a GRW-type collapse mechanism.
In their model, however, the particle (or beable) in no way affects the collapse mechanism.
There is, therefore, no action-reaction principle between wave and particle.
In our model, though, the evolution of the wavefunction is directly affected by the particle position.  
For the following we will limit ourselves to nonrelativistic, closed systems subject to a time-independent Hamiltonian.
We will also assume, for simplicity, that the system has a discrete energy spectrum.
We begin with the single particle case, then extend the formalism to the many-particle case.

\subsection{Basic principles}
\label{Sb_BP}
We begin by stating the basic principles of the dualist interpretation.~\\

1 - \emph{Ontology}: The dualist interpretation has the same ontology as pilot-wave interpretations.
There is a wavefunction $\psi$, which is considered objectively real.
In addition, there is a particle, considered real, with a well defined trajectory, ${\bf x}_{p} = {\bf x}_{p}(t)$.~\\

2 - \emph{The Wave-to-Particle (W-P) Guidance Condition}: The movement of the particle is described by a stochastic process controlled by the wavefunction.
This stochastic process is identical to the one in the SPW interpretation, and is described by the Langevin equation (\ref{E_Lgvn}).~\\

3 - \emph{The Particle-to-Wave (P-W) Guidance Condition}: The P-W guidance condition is modeled by successive, discontinuous changes in the wavefunction, called \emph{Spontaneous Transition Events} (STE).
The STE are Poisson distributed in time with a frequency parameter $\lambda$.
The new wavefunction is chosen from a set of accessible wavefunctions, defined as the set of wavefunctions that conserve certain values of the wavefunction prior to the STE.
Here, we postulate that every conserved quantity of the wavefunction, with respect to the Schr\"{o}dinger equation, is stricly conserved with respect to a STE.
We call this the \emph{Maximal Strict Conservation} (MSC) principle.
We will discuss possible alternatives in Section \ref{S_Disc}.
Also, the probability of obtaining a particular member of the accessible set after a STE is proportional to the squared amplitude of that normalized wavefunction at the particle position.
In that way, the particle guides the transition. 

\subsection{The single particle}
\label{Sb_SP}
\subsubsection{Formulation}
\label{Ssb_For1}
The wavefunction of a single particle is expanded into its energy eigenstates:
\begin{equation}
\psi({\bf x},t) = \sum_{i = 0}^{K} \sum_{j = 0}^{M_{i}-1} c_{i j} \phi_{i j}({\bf x}\,) \exp[-i E_{i} t / \hbar]
\label{E_EEG}
\end{equation}
where the index $i$ numbers the $K+1$ energy eigenvalues with a non-zero amplitude, the index $j$ numbers the $M_{i}$ degenerate eigenstates with energy $E_{i}$, and $c_{i j}$ are the complex coefficients for each eigenstate $\phi_{i j}({\bf x})$.
It is convenient to rewrite Eq. (\ref{E_EEG}) as
\begin{equation}
\psi({\bf x},t) = \sum_{i = 0}^{K} C_{i} \Phi_{i}({\bf x}\,) \exp[-i E_{i} t / \hbar]
\label{E_EEG2}
\end{equation}
where we have defined
\begin{mathletters}
\label{E_Ddf}
\begin{equation}
\Phi_{i}({\bf x}) = C_{i}^{-1} \sum_{j = 0}^{M_{i}-1} c_{i j} \phi_{i j}({\bf x}\,)
\label{E_Ddfa}
\end{equation}
\begin{equation}
C_{i} = \Bigg{(} \sum_{j = 0}^{M_{i}-1} | c_{i j} |^{2} \Bigg{)}^{\frac{1}{2}} \exp[i \, \Theta_{i}] {\rm .}
\label{E_Ddfb}
\end{equation}
\end{mathletters}
Since the phases $\Theta_{i}$ are arbitrary, we set $\Theta_{i} = \theta_{i 0}$, where $\theta_{i j}$ is the phase of $c_{i j}$.

\subsubsection{Spontaneous transitions}
\label{Ssb_ST}
By virtue of the MSC principle, only the phases $\Theta_{i}$ change as a result of a STE.
All the other quantities, $|C_{i}|$ and $\Phi_{i}({\bf x})$, do not change.
The constants of the motion, therefore, stay constant with respect to a transition.
In what follows, we will express the wavefunction as a function of these phases, $\psi = \psi({\bf x},\vec{\Theta},t)$, where $\vec{\Theta} = (0,\Theta_{1},\Theta_{2},...,\Theta_{K})$ and where we keep $\Theta_{0} = 0$ to eliminate an arbitrary overall phase factor.

If a STE occurs at time $t_{0}$, where the initial wavefunction and particle position are $\psi({\bf x},\vec{\Theta},t_{0})$ and ${\bf x}_{p}(t_{0})$, respectively, then the wavefunction undergoes the transition $\psi({\bf x},\vec{\Theta},t_{0}) \rightarrow \psi({\bf x},\vec{\Theta}',t_{0})$, where $\vec{\Theta}'$ is the new phase vector.
The particle position, however, does not change because of the STE.
As mentioned previously, the conditional probability density of the new phase vector, $f(\vec{\Theta}'|{\bf x}_{p},t_{0})$, is proportional to the squared amplitude of the (normalized) wavefunction at the particle position,
\begin{equation}
f(\vec{\Theta}'|{\bf x}_{p},t_{0}) = \Gamma^{-1} |\psi({\bf x}_{p},\vec{\Theta}',t_{0})|^{2}
\label{E_STE}
\end{equation}
where $\Gamma = \Gamma({\bf x}_{p})$ is a normalizing factor.
\begin{mathletters}
\label{E_T}
\begin{equation}
\Gamma({\bf x}_{p}) = \int_{0}^{2 \pi}...\int_{0}^{2 \pi} |\psi({\bf x}_{p},\vec{\Theta}',t_{0})|^{2} \prod_{i=1}^{K} {\rm d} \Theta_{i}'
\label{E_Ta}
\end{equation}
\begin{equation}
\Gamma({\bf x}_{p}) = (2 \pi)^{K} \sum_{i = 0}^{K} |C_{i}|^{2} |\Phi_{i}({\bf x}_{p})|^{2}
\label{E_Tb}
\end{equation}
\end{mathletters}
The integration over the phase vector in Eq. (\ref{E_Ta}) eliminates the interference terms between energy eigenstates, which leads to Eq. (\ref{E_Tb}) and implies that $\Gamma$ depends only on the particle position.

In Eq. (\ref{E_STE}), the squared amplitude is no longer seen as the conditional probability density of the particle position for a given wavefunction, but rather it is the conditional probability density of the wavefunction for a given particle position.
Born's statistical postulate has been turned on its head, thereby giving the squared amplitude a dual meaning.
In the language of statistical inference, we would say that $f(\vec{\Theta}'|{\bf x}_{p},t_{0})$ is proportional to the likelyhood that the probability density which caused the observation ${\bf x}_{p}$ at time $t_{0}$, had the parameter values $\vec{\Theta}'$.
In this case, however, Eq. (\ref{E_STE}) represents a real stochastic process of the wavefunction, conditioned by the particle.
Unlike the GRW model, there is no collapse as such, only a random change in the phases which favors wavefunctions that are peaked and centered about the particle.
Although an expansion of the wavefunction as a result of a STE is not impossible.
In other words, the STE is a `quantum jump' from one solution of the Schr\"{o}dinger equation to another which better represents, on average, the current particle position (while staying within the set of accessible states).

Given an initial probability density for the particle position, $\rho({\bf x}_{p},t_{0})$, the marginal probability density of the new phase vector is
\begin{equation}
f(\vec{\Theta}',t_{0}) = \int f(\vec{\Theta}'|{\bf x}_{p},t_{0}) \rho({\bf x}_{p},t_{0}) {\rm d}^{3} x_{p} {\rm .}
\label{E_STEF1}
\end{equation}
If quantum equilibrium holds prior to the transition, Eq. (\ref{E_STEF1}) equals
\begin{equation}
f(\vec{\Theta}'|\vec{\Theta},t_{0}) = \int f(\vec{\Theta}'|{\bf x}_{p},t_{0}) |\psi({\bf x}_{p},\vec{\Theta},t_{0})|^{2} {\rm d}^{3} x_{p}
\label{E_STEF2}
\end{equation}
which is the transition probability density function for a STE.
We can also use Bayes rule to determine the particle position probability density after the STE for a given final wavefunction $\vec{\Theta}'$, $\rho '({\bf x}_{p}|\vec{\Theta}',t_{0})$,
\begin{equation}
\rho '({\bf x}_{p}|\vec{\Theta}',t_{0}) = \frac{\rho({\bf x}_{p},t_{0}) f(\vec{\Theta}'|{\bf x}_{p},t_{0})}{f(\vec{\Theta}',t_{0})}{\rm .}
\label{E_rho2}
\end{equation}
Again, if quantum equilibrium obtained prior to the STE, we can derive from Eq. (\ref{E_rho2}) the expression,
\begin{equation}
\rho '({\bf x}_{p}|\vec{\Theta}',t_{0}) = \bigg{[}\frac{f(\vec{\Theta}|{\bf x}_{p},t_{0})}{f(\vec{\Theta}'|\vec{\Theta},t_{0})}\bigg{]} |\psi({\bf x}_{p},\vec{\Theta}',t_{0})|^{2} 
\label{E_rho2b}
\end{equation}
where we have made use of Eqs. (\ref{E_STE}) and (\ref{E_STEF1}).
Clearly, Eq. (\ref{E_rho2b}) shows that quantum equilibrium is not necessarily conserved after a transition event.
Quantum equilibrium may be a good approximation if the initial wavefunction is much wider than the final one, since the initial wavefunction would be essentially constant over the region where the final wavefunction is appreciable.
Such a situation may occur if, in between two consecutive STEs, the wavefunction has enough time to spread out in position space, and that the STE yields a reasonably narrow final wavefunction.
Whatever deviations from quantum equilibrium may arise as a result of the STE will eventually vanish due to the stochastic evolution of the particle.
Quantum equilibrium would be a good approximation for particle statistics if the timescale for it is much smaller than the average lifetime between transitions $\lambda^{-1}$.

Furthermore, should the system possess a well-defined energy, $K = 0$, then the transition events would have no effect on it.
In the double-slit experiment, for instance, if we have an initial plane wave with a well-defined energy, then the two branches of the wavefunction that are produced by the slits would be have the same energy.
The coherence between them would be preserved by a STE.
Therefore, even if a transition event were to occur while the particle is travelling between the slits and the screen, it would have no effect on the interference pattern.

Equations (\ref{E_STEF1}) to (\ref{E_rho2b}) only apply to the case where the initial state is pure.
But for an initial mixed state, the results can be quite different.
If we have an initial ensemble of states that all belong to the same set of accessible states, but is completely decoherent (i.e. every component of the phase vector is independent and uniformly distributed between 0 and $2\pi$), and assuming quantum equilibrium for every member of the ensemble, then the initial particle position probability density is $\rho ({\bf x}_{p}) = (2\pi)^{-K} \Gamma ({\bf x}_{p})$.
Placing this result in Eq. (\ref{E_STEF1}), the marginal probability density of the new phase vector is $f(\vec{\Theta}') = (2\pi)^{-K}$, which is the probability for that phase vector in the initial ensemble.
Furthermore, placing these results in Eq. (\ref{E_rho2b}), it is easily shown that the particle position probability density after the STE for the final wavefunction $\vec{\Theta}'$ is given by $\rho '({\bf x}_{p}|\vec{\Theta}',t_{0}) = |\psi({\bf x}_{p},\vec{\Theta}',t_{0})|^{2}$.
Therefore, not only is the probability distribution of the wavefunction for this mixed state conserved after a transition event, but so is the quantum equilibrium for every wavefunction in the ensemble.
And since neither the Schr\"{o}dinger equation nor the W-P guidance condition can alter this joint wave-particle probability distribution, this mixed state is therefore completely stationary.
Owing to its features of complete decoherence and quantum equilibrium, we shall call this mixed state the \emph{Decoherent Quantum Equilibrium} (DQE) steady state.
It must be emphasized that the term `decoherence' in the dualist interpretation simply means that the relative phases are random.
It does not means that the phases don't exist, or that they are undecidable, or that a wavefunction collapse took place.

The DQE steady state has two desirable properties.
First, since it is a steady state, then any initial pure state, in quantum equilibrium or not, will irreversibly evolve towards its corresponding DQE mixed state and will stay in that state.
This introduces a kind of time's arrow in the dualist interpretation.
Second, the canonical ensemble of a system in thermal equilibrium can always be decomposed into a combination of DQE steady states.
Therefore, any microscopic system that is extracted from this ensemble must be in quantum equilibrium, regardless of any STEs that may have occured prior to the extraction.
This guarantees that the results of orthodox quantum theory are reproduced in experiments like the one by Arndt et al. \cite{Arndt.et.al99}, where C$_{60}$ molecules are sublimated from a 900-1000 K oven to a diffraction grating which causes an interference pattern.

\subsubsection{General stochastic evolution}
An initial pure state will evolve into a mixed state as a result of the spontaneous transition events.
Therefore, given an initial pure state, $\psi({\bf x},\vec{\Theta},t_{0})$, and particle position, ${\bf x}_{p}(t_{0})$, the  position expectation value of the wavefunction will evolve as,
\begin{mathletters}
\label{E_mu}
\begin{equation}
\{ \vec{\mu}(t_{0}+{\rm d}t)\} = (1-\lambda {\rm d}t)[\,\vec{\mu}(t_{0}) + \frac{\langle {\bf p} \, \rangle}{m} {\rm d}t] + \{\vec{\mu}(t_{0})\}_{p} \lambda {\rm d}t
\label{E_mua}
\end{equation}
\begin{equation}
\{\vec{\mu}(t_{0}+{\rm d}t)\} = \vec{\mu}(t_{0}) + \frac{\langle {\bf p}\, \rangle}{m} {\rm d}t + [\,\{\vec{\mu}(t_{0})\}_{p} - \vec{\mu}(t_{0})] \lambda {\rm d}t
\label{E_mub}
\end{equation}
\end{mathletters}
where $\vec{\mu} = \langle \psi|\vec{x} \, |\psi \rangle$ is the position expectation value of the wavefunction, $\langle {\bf p} \,\rangle = \langle \psi|{\bf p}\,|\psi \rangle$ is the momentum expectation value of the wavefunction, and the curly brackets indicate an ensemble average in a mixed state.
It is important to note that, in this context, $\vec{\mu}$ is the centroid of the wavefunction (i.e. a geometrical point representing the wavefunction), and not the average particle position.
Therefore, 
\begin{equation}
\{\vec{\mu}(t)\}_{p} = \int \langle \psi({\bf x},\vec{\Theta}',t)|{\bf x} \, |\psi({\bf x},\vec{\Theta}',t) \rangle f(\vec{\Theta}'|{\bf x}_{p},t) {\rm d}^{K} \Theta'
\label{E_cup}
\end{equation}
is the average position of the mixed state created by the STE.
Also, we have neglected the term proportional to $({\rm d}t)^{2}$ in Eq. (\ref{E_mub}).
If we assume that the ensemble of wavefunctions created by the STE is symmetrically distributed about the particle position, then $\{\vec{\mu}\}_{p} = {\bf x}_{p}$ and Eq. (\ref{E_mub}) becomes,
\begin{equation}
\{\vec{\mu}(t_{0}+{\rm d}t)\} = \vec{\mu}(t_{0}) + \frac{\langle {\bf p} \, \rangle}{m} {\rm d}t - [\,\vec{\mu}(t_{0}) - {\bf x}_{p}(t_{0})] \lambda {\rm d} t
\label{E_mu2}
\end{equation}
where the square bracket term attracts, in a sense, the average position of the wavefunctions towards the particle.
In this way, we may interpret the influence of the particle as a \emph{pilot-particle} guiding the evolution of the wavefunction, which, in turn, guides the evolution of the particle.
Similarly, the variance of the wavefunction, $\sigma^{2} = \langle \psi|({\bf x}-\langle {\bf x} \rangle)^{2}|\psi \rangle$, evolves as,
\begin{equation}
\{\sigma^{2}(t_{0}+{\rm d}t)\} = \sigma^{2}(t_{0}) + \chi (t_{0}) {\rm d}t - [ \, \sigma^{2}(t_{0}) - \{\sigma^{2}(t_{0})\}_{p}] \lambda {\rm d} t
\label{E_sige}
\end{equation}
where $\chi = {\rm d} \sigma^{2} / {\rm d} t$ is the rate of growth of the variance of the wavefunction described by the Schr\"{o}dinger equation, and $\{\sigma^{2}\}_{p}$ is the average variance of the mixed state created by the STE.
The particle, therefore, can be seen to impede the growth of the (average) variance of the wavefunction whenever $\sigma^{2} > \{\sigma^{2}\}_{p}$.

\subsection{The many-particle system}
\label{Sb_MPE}
\subsubsection{Formulation}
\label{Ssb_For2}
We now extend the dualist interpretation to the case with an arbitrary number of particles, $N$.
It is convenient, therefore, to introduce the notation, $\vec{q}_{p} = ({\bf x}_{p1},{\bf x}_{p2},...,{\bf x}_{pN})$, where $\vec{q}_{p}$ is the vector of particle positions in the configuration space $\vec{q} = ({\bf x}_{1},{\bf x}_{2},...,{\bf x}_{N})$.
The wavefunction representation in Eq. (\ref{E_EEG2}) has the straightforward extension:
\begin{equation}
\psi(\vec{q},t) = \sum_{i=0}^{K} C_{i} \Phi(\vec{q}\,) \exp[-i E_{i} t / \hbar]
\label{E_EEGN}
\end{equation}
where the $\Phi(\vec{q}\,)$ are coherent superpositions of degenerate energy eigenstates for the many-particle system.
Here, we assume that the wavefunction is non-separable, which means that at least one of the following two conditions must apply.
First, it is impossible to express the wavefunction as a product of two wavefunctions: $\psi(\vec{q},t) \ne \psi(\vec{q}_{n},t) \psi(\vec{q}_{(N-n)},t)$ for all time, where $\vec{q}_{n}$ is the configuration space vector of a subset of $n$ particles and $\vec{q}_{(N-n)}$ is the complement configuration space vector of the remaining $N-n$ particles.
In other words, the particles must be entangled.
Second, the particles must interact with one another.
A separable wavefunction, therefore, is one where the wavefunction may be factorized, for all time, into two or more wavefunctions of subsets of particles where the particles in one subset do not interact with the particles in another subset.

From Eq. (\ref{E_EEGN}), it follows that the extension of Eq. (\ref{E_STE}) to a many-particle system is,
\begin{equation}
f(\vec{\Theta}'|\vec{q}_{p},t_{0}) = \Gamma^{-1} |\psi(\vec{q}_{p},\vec{\Theta}',t_{0})|^{2}
\label{E_STEN}
\end{equation}
where $\Gamma = \Gamma(\vec{q}_{p})$ is a normalizing factor similar to the one used in the single-particle case, and is equal to,
\begin{equation}
\Gamma(\vec{q}_{p}) = (2 \pi)^{K} \sum_{i=0}^{K} |C_{i}|^{2} |\Phi(\vec{q}_{p})|^2 {\rm .}
\label{E_TN}
\end{equation}
We postulate that the STEs described by Eq. (\ref{E_STEN}) are Poisson distributed in time with a frequency parameter $N \lambda$.
This is because any one particle may `trigger' a STE with a frequency $\lambda$.
But since all the particles are entangled, they must all condition the transition event.
For any given STE, therefore, every particle is equally likely to have triggered it, independently of every other particle.
The frequencies must then add to $N \lambda$.
Consequently, if the wavefunction is separable into the subsets $\vec{q}_{pn} = ({\bf x}_{p1},...,{\bf x}_{pn})$ and $\vec{q}_{p(N-n)} = ({\bf x}_{p(N-n)},...,{\bf x}_{pN})$, for instance, then the phase vector of the whole wavefunction $\vec{\Theta}$ can be expressed as two phase vectors: $\vec{\Theta}_{n}$ which is guided only by $\vec{q}_{pn}$, and $\vec{\Theta}_{N-n}$ which is guided only by $\vec{q}_{p(N-n)}$.
The wavefunction represented by $\vec{\Theta}_{n}$ undergoes a transition with a frequency $n \lambda$, and the $\vec{\Theta}_{N-n}$ wavefunction with a frequency $(N-n) \lambda$, independently of the transitions for $\vec{\Theta}_{n}$.

\subsubsection{The macroscopic limit}
Now we will examine a simplified account of the motion of a body with a large number of particles, $N \approx 10^{23}$, held together by an interparticle potential, $V_{nm}(|{\bf x}_{n}-{\bf x}_{m}|)$.
We assume that the wavefunction of this body can be factorized as,
\begin{equation}
\psi(\vec{q},t) = \Psi({\bf R},t) \zeta(\Delta \vec{q},t)
\label{E_cm}
\end{equation}
where ${\bf R} = \sum_{n=1}^{N} m_{n} {\bf x}_{n} / (\sum_{n=1}^{N} m_{n})$ is the center-of-mass coordinate of the body and where $\Delta \vec{q}$ is the vector of all interparticle distances, ${\bf r}_{nm} = {\bf x}_{n} - {\bf x}_{m}$, for $n < m$.
We also assume that the wavefunction of the internal variables of the body is stationary: $\zeta(\Delta \vec{q}, t) \rightarrow \zeta(\Delta \vec{q}\,) \exp [-i E_{{\rm int}} t / \hbar]$, where $E_{{\rm int}}$ is the internal energy of the body.
The center-of-mass wavefunction, however, is not stationary,
\begin{equation}
\Psi({\bf R},t) = \sum_{i = 0}^{K} C_{i} \Phi_{i}({\bf R} \,) \exp [-i E_{i} t / \hbar]{\rm .}
\label{E_cmwf}
\end{equation}
This allows us to attribute a phase vector, $\vec{\Theta} = (\Theta_{1},...,\Theta_{K})$, to the center-of-mass wavefunction, $\Psi({\bf R},t) \rightarrow \Psi({\bf R},\vec{\Theta},t)$, but not to the internal wavefunction (since $K = 0$ in that case).
Placing the wavefunction described in Eq. (\ref{E_cm}) in the STE equation (\ref{E_STEN}), we obtain
\begin{equation}
f(\vec{\Theta}'|{\bf R}_{p},t_{0}) = \Gamma^{-1} |\Psi({\bf R}_{p},\vec{\Theta}',t_{0})|^{2}
\label{E_STER}
\end{equation}
where ${\bf R}_{p} = \sum_{n=1}^{N} m_{n} {\bf x}_{pn} / (\sum_{n=1}^{N} m_{n})$ is the center-of-mass of the particle positions, and 
\begin{equation}
\Gamma({\bf R}_{p}) = (2 \pi)^{K} \sum_{i = 0}^{K} |C_{i}|^{2} \Phi_{i}({\bf R}_{p})
\label{E_TR}
\end{equation}
is the normalizing factor.
The internal wavefunction does not appear in Eq. (\ref{E_STER}) because it is independent of $\vec{\Theta}'$, and so would cancel itself out upon renormalization.
The STEs, therefore, are guided solely by the center-of-mass ${\bf R}_{p}$, and since all $N$ particles are entangled, the STEs occur with a frequency parameter $N \lambda$.
Here, we have managed to reproduce one of the results of the GRW theory, namely, that for a similar system the hits only affect the center-of-mass coordinate and occur with a frequency $N \lambda$.

For a macroscopic object, then, the time between transition events can be very much shorter than for a microscopic object.
The expectation value equations for a single particle, (\ref{E_mu2}) and (\ref{E_sige}), can now be extended to the many-particle case,
\begin{mathletters}
\label{E_muR}
\begin{equation}
\{\vec{\mu}(t_{0}+{\rm d}t)\} = \vec{\mu}(t_{0}) + \frac{\langle {\bf P} \, \rangle}{M} {\rm d}t - [\,\vec{\mu}(t_{0}) - {\bf R}_{p}(t_{0})] (N \lambda) {\rm d} t
\label{E_muRa}
\end{equation}
\begin{equation}
\{\sigma^{2}(t_{0}+{\rm d}t)\} = \sigma^{2}(t_{0}) + \chi (t_{0}) {\rm d}t - [ \, \sigma^{2}(t_{0}) - \{\sigma^{2}(t_{0})\}_{Rp}] (N \lambda) {\rm d} t
\label{E_muRb}
\end{equation}
\end{mathletters}
where $\vec{\mu} = \langle \Psi | {\bf R} \, | \Psi \rangle$, $\langle {\bf P} \,\rangle = \langle \Psi | {\bf P} \, | \Psi \rangle$, ${\bf P}$ being the center-of-mass momentum, $M = \sum_{n=1}^{N} m_{n}$ is the total mass, $\sigma^{2} = \langle \Psi | ({\bf R} - \langle {\bf R} \, \rangle)^{2} | \Psi \rangle$, and where $\{\sigma^{2}\}_{Rp}$ is analogous to $\{\sigma^{2}\}_{p}$  in Eq. (\ref{E_sige}).
Because the frequency parameter for the center-of-mass STEs of a macroscopic object is much greater than that of a microscopic object, the influence of the center-of-mass on its wavefunction is proportionately greater.
Therefore, the influence of the center-of-mass on the position and variance expectation values of its wavefunction is greatly increased.

The Langevin equation for the center-of-mass, ${\bf R}_{p}$, is similar to the single particle equation (\ref{E_Lgvn}).
\begin{equation}
{\bf R}_{p}(t+{\rm d}t) = {\bf R}_{p}(t) + {\bf b}({\bf R}_{p},t){\rm d}t + D^{1/2} {\bf w}(t) ({\rm d}t)^{1/2}
\label{E_LgvnR}
\end{equation}
where $D = \hbar / M$.
The drift function is given by
\begin{equation}
{\bf b} = \frac{{\bf \nabla}_{R} {\cal S}}{M} + \frac{\hbar}{M}\frac{{\bf \nabla}_{R} {\cal R}}{{\cal R}}
\label{E_drftR}
\end{equation}
where we set $\Psi = {\cal R} \exp[i {\cal S}/\hbar]$.
In order to analyze the macroscopic limit, we look at the center-of-mass equations for a body that starts with $N = 1$, and we progressively add particles to the body until $N \gg 1$.
We will assume, for simplicity, that all the particles have an equal mass, so that $M = Nm$.
It is convenient to introduce a quantum equilibrium timescale $ \tau $, such that $\rho({\bf R}_{p},t_{0}+ \tau | {\bf R}_{p0},t_{0}) \approx P({\bf R}_{p},t_{0}+ \tau )$, which scales as,
\begin{equation}
\tau(N) = \frac{L^{2}(N)}{D} = \frac{Nm}{\hbar}L^{2}(N)
\label{E_tau}
\end{equation}
where $L(N)$ is the characteristic length scale the squared amplitude of the wavefunction.
Its exact value is not important, all that matters is how it changes with increasing $N$.
Note that the approach towards quantum equilibrium depends not only on the stochastic diffusion, controlled by $D$, but also on the mixing by the wavefunction, controlled by the drift function ${\bf b}$.
However, we will ignore mixing effects and assume that diffusion controls the rate at which quantum equilibrium is approached.
If we assume the particles are statistically independent and that their respective wavefunctions are separable, then the length scale should change like the standard deviation of the average of $N$ independent random variables, $L(N) = N^{-1/2} L(1)$.
This would mean that the timescale scales as $\tau(N) = m L(1) / \hbar = \tau(1)$, i.e. that $\tau$ does not change with $N$.
If we assume that $L(1) \approx 10^{-5} {\rm cm}$, then for a proton, $\tau(1) \approx 10^{-7} {\rm s}$, while for an electron $\tau(1) \approx 10^{-11} {\rm s}$.
The particles are not really independent, of course, since they are bound together to form a body.
Nevertheless, as we will see, the conclusions drawn from this section do not depend critically of the scaling for $L$.

For the first time, we will impose a constraint on the value of the frequency parameter, $\lambda$.
We require that $\lambda \tau(1) \ll 1$, meaning that microscopic objects would have ample time to reach quantum equilibrium before the next STE which might break it.
If we adopt the same value for the frequency parameter as in the GRW model, then this condition is certainly satisfied, where $\lambda \tau(1) \approx 10^{-23}$ for a proton.
For a macroscopic object with $N \approx 10^{23}$, then $N \lambda \tau(1) \approx 1$, meaning that the time to reach quantum equilibrium is, on average, the same as the time between successive STEs.
However, it is conceivable to make $N$ large enough so that $N \lambda \tau(1) \gg 1$.
In that case, the STEs occur too frequently for quantum equilibrium to be valid in general.
In that limit, the conditional particle position variance, $\sigma^{2}(t_{0}+\Delta t | {\bf R}_{p0},t_{0})$ (where $t_{0}$ is the time of the previous STE), has not had enough time to spread appreciably before the following STE,
\begin{equation}
\sigma^{2}(t_{0} + (N \lambda)^{-1} | {\bf R}_{p0},t_{0}) \approx \hbar / (N^{2} m \lambda) \ll L^{2}(1) / N {\rm .}
\label{E_sigdel}
\end{equation}

If we further assume that the drift function does not vary appreciably between two STEs, and over the particle displacement, then we can say that the conditional particle position average is approximately,
\begin{equation}
\{ {\bf R}_{p} \} (t_{0} + (N \lambda )^{-1} | {\bf R}_{p0}, \vec{\Theta}', t_{0}) \approx {\bf R}_{p0}(t_{0}) + {\bf b}({\bf R}_{p0},\vec{\Theta}',t_{0}) (N \lambda)^{-1}
\label{E_AVER}
\end{equation} 
where we have placed the wavefunction phase vector $\vec{\Theta}'$ in the arguments of the conditional average and the drift function to explicitly show their dependence on the state of the wavefunction.
However, if we only know the initial center-of-mass position ${\bf R}_{p0}(t_{0})$, and we do not know \emph{a priori} which phase vector was chosen by the STE, then we must take the average with respect to the conditional probability density $f(\vec{\Theta}'|{\bf R}_{p0},t_{0})$.

Before we can find that average, it is convenient to write the drift function as,
\begin{equation}
{\bf b} = \frac{\hbar}{\sqrt{2} M |\Psi|^{2}}( e^{-i (\pi /4)} \Psi^{*} {\bf \nabla}_{R} \Psi + e^{i (\pi /4)} \Psi {\bf \nabla}_{R} \Psi^{*} ) {\rm .}
\label{E_drftb}
\end{equation}
Combining Eqs. (\ref{E_cmwf}) and (\ref{E_STER}) with Eq. (\ref{E_drftb}), and averaging over $\vec{\Theta}'$, we obtain,
\begin{equation}
\{ {\bf b}({\bf R}_{p0},t_{0}) \}_{Rp} = \frac{1}{\Gamma({\bf R}_{p0})} \sum_{i = 0}^{K} |C_{i}|^{2} |\Phi({\bf R}_{p0})|^{2} {\bf b}_{i}({\bf R}_{p0})
\label{E_adft}
\end{equation}
where ${\bf b}_{i}$ is the drift function corresponding to the stationary state $\Phi_{i}$,
\begin{equation}
{\bf b}_{i} = \frac{\hbar}{\sqrt{2}M |\Phi_{i}|^{2}}(e^{-i(\pi /4)} \Phi^{*}_{i} {\bf \nabla}_{R} \Phi_{i} + e^{i(\pi /4)} \Phi {\bf \nabla}_{R} \Phi^{*}_{i}){\rm .}
\label{E_dfti}
\end{equation}
We can gain more insight into these equations by considering the simple case of a free body in one dimension.
Specifically, we consider a Gaussian wavepacket centered at $R = 0$, with a standard deviation $\sigma$ at $t = 0$, and with a mean velocity $U$,
\begin{equation}
\Psi(R,t) = \bigg{[}\frac{\sigma}{\sqrt{2 \pi^{3}} \hbar^{2}}\bigg{]}^{\frac{1}{2}} \int_{0}^{\infty} \big{[}e^{-\sigma^{2}(P - MU)^{2}/\hbar^{2} + iPR/\hbar} + e^{-\sigma^{2}(P + MU)^{2}/\hbar^{2} - iPR/\hbar}\big{]} \, e^{-i P^{2}t/(2M\hbar)} {\rm d} P
\label{E_Gwf}
\end{equation}
where the square brackets in the integral include the two degenerate eigenstates for each energy level.
Instead of a phase vector, we now have a continuous phase function $\Theta(P)$, corresponding to the continuous energy spectrum.
By adapting Eq. (\ref{E_adft}) for the continuous energy spectrum case, we obtain
\begin{equation}
\{ b(R_{p}) \}_{Rp} = \frac{U - [\hbar R_{p}/(2M\sigma^{2})]e^{-2(\sigma M U /\hbar)^{2} - R_{p}^{2}/(2 \sigma^{2})}}{1+ e^{-2(\sigma M U /\hbar)^{2}-R_{p}^{2}/(2 \sigma^{2})}}{\rm .}
\label{E_Gadrft}
\end{equation}
The average drift in Eq. (\ref{E_Gadrft}), like the average drift described in Eq. (\ref{E_adft}), is time-independent.
This is because it has been evaluated from a spectrum of stationary states.
Also, as $|R_{p}/\sigma| \rightarrow \infty$, the average drift approaches the mean velocity, $\{ b(R_{p}) \}_{Rp} \rightarrow U$.
In the region $|R_{p}/\sigma| \le 1$, the average drift may deviate from the mean velocity by as much as $|\{ b(R_{p}) \}_{Rp} - U| \leq \hbar /(2M\sigma)$.
The deviation in the region $|R_{p}/\sigma| \leq 1$ is a consequence of the preservation of the relative phases between energy eigenfunctions in a STE.
As we have already assumed, the standard deviation of the wavefunction scales as $\sigma \propto N^{-1/2}$, which means that the region over which the average drift deviates from the mean velocity shrinks with increasing $N$.
Also, since $M = N m$, the deviation scales as $N^{-1/2}$.
Consequently, for very large bodies, the average drift is equal to $U$ to a very good approximation.
The average of Eq. (\ref{E_AVER}) with respect to a STE, and in the limit of very large $N$, tends to
\begin{equation}
\{ R_{p} \} (t_{0} + (N \lambda )^{-1} | R_{p0}, t_{0}) \approx R_{p0}(t_{0}) + U (N \lambda)^{-1}{\rm .}
\label{E_GAVER}
\end{equation}
Physically, Eq. (\ref{E_GAVER}) states that for very large bodies, the particles, represented here by the center-of-mass coordinate, dominate the evolution of the center-of-mass wavefunction in such a way that the osmotic velocity of the drift function averages out over many STEs.
Likewise, the particle velocity $\partial_{R} {\cal S}/M$ averages to the mean velocity $U$, which is a constant and does not change as a result of the STEs.
The center-of-mass not only `drags' the wavefunction along, as in Eq. (\ref{E_muRa}), but, on average, also assumes the overall velocity of the wavefunction.

There remains the question of the variance of the center-of-mass position at the time $t_{0} + (N \lambda)^{-1}$.
This variance is the sum of a diffusion component, $D/(N \lambda)$, and a component due to the variance of the drift function, $\sigma^{2}_{b}/(N \lambda)$.
Given that the diffusion constant $D$ scales as $N^{-1}$, we may neglect this component for very large bodies.
We will not attempt to evaluate $\sigma^{2}_{b}$ here, but it reasonable to assume that it is a function of the velocity uncertainty of the wavefunction, a constant for this system, and weakly dependent on $R_{p}/\sigma$, in the same way and for the same reason as for the average drift.

Here we have shown, in a limited way, how in the macroscopic limit the particle aspect dominates the dynamics of the wave-particle system, while assuming, on average, certain characteristics of the wavefunction.
The principal assumption used in this development is the scaling law of the characteristic length of the squared amplitude of the wavefunction, $L \propto N^{-1/2}$.
This assumption, while arbitrary, is not crucial.
If $L$ were constant (which may be the case if the particles were perfectly correlated with each other), or increased with respect to the particle number, then the limit $N \lambda \tau(N) \gg 1$ would be attained faster with respect to $N$ than in the case presented here.
Also, if $L$ were constant, for instance, then the region over which the average drift in Eq. (\ref{E_GAVER}) deviates from $U$ would also be constant, but the magnitude of the deviation would scale as $N^{-1}$.
If, on the other hand, $L$ decreased with increasing $N$ such that $L \propto N^{-\chi}$ and $\chi > 1/2$, then the macroscopic limit may never be attained.
However, these scaling laws only make sense in the context of a specific preparation procedure for the macroscopic body.
It is this procedure that determines the change in $L$ as new particles are incorporated into the body.
It is important to note that, while limited, this demonstration shows that in the macroscopic limit the dualist interpretation provides an approximately classical trajectory for a closed system, without the need for environmental decoherence.
In this sense, macroscopic bodies are self-decohering.

\section{Discussion}
\label{S_Disc}

\subsection{Comparison of interpretations}
Bell, in Ref. \cite{Bell90}, stated that the resolution of the measurement problem of orthodox quantum mechanics would come \emph{either} from a Bohm-type pilot-wave interpretation, \emph{or} a GRW-type wavefunction collapse interpretation.
We have attempted, however, to formulate an interpretation where the \emph{or} in that statement is replaced with an \emph{and}, although we don't have, strictly speaking, a GRW-type collapse mechanism.
Rather, we have a stochastic process where the wavefunction undergoes random transitions that are conditioned, or guided, by the particle.

At first sight, the dual ontology coupled with the dual dynamics of the wavefunction may seem unduly complicated.
Proponents of pilot-wave interpretations may question the necessity of introducing a stochastic wavefunction process in addition to the deterministic Schr\"{o}dinger equation.
After all, the introduction of the particle already solves the measurement problem without the need for an additional wavefunction process.
Likewise, proponents of an explicit collapse interpretation might wonder why the introduction of a particle is needed, since the explicit collapse mechanism already accounts for the determinateness of the macroscopic world.

In reponse to the first objection, we would like to point out that in the dualist interpretation, the wave and the particle are on a more equal footing than in the pilot-wave interpretations.
Here, the particle is an active participant in the evolution of the wave-particle system.
Indeed, a dual guidance condition is a natural extension of pilot-wave interpretation: since both wave and particle are real, and the wave guides the particle, then the particle should also guide the wave.
Consequently, the particle cannot be dismissed as an \emph{ad hoc} device introduced into the mathematical formalism of quantum mechanics for the sole purpose of getting rid of the measurement problem.
Furthermore, the stochastic action of the particle on the wavefunction leads to an \emph{objective} decoherence process necessary to eliminate empty wave effects and to ensure the emergence of classical trajectories in the macroscopic limit.
The term `objective' emphasizes the fact that the particle-guided STEs introduce an irreducible randomization of relative phases in the energy basis.
The decoherence is not a consequence of our ignorance of a complicated but deterministic external environment acting on the system.
Indeed, the model described here refers to closed systems, which eventually become completely decoherent in the energy basis on their own.
Environmental decoherence not only still applies to open systems, but now possesses a stronger theoretical foundation within the dualist interpretation.

The role of the wavefunction can be seen, therefore, to generate a quantum equilibrium ensemble for the particle, while the particle generates an energy decoherent ensemble for the wavefunction, together they tend towards a DQE steady state.
Degenerate energy eigenstates, however, retain their coherence in the energy decoherent ensemble generated by the particle in a closed system.
On the other hand, it is reasonable to assume that for open systems, the combination of particle-guided STEs and the interaction with the environment would generate a completely decoherent wavefunction ensemble for the subsystem under consideration.
The interaction with the environment would break the degeneracy of the closed-system eigenstates, and the STEs would, over time, generate a genuinely random wavefunction ensemble.
In that sense, the action of the particles on the wavefunction provides a link between quantum mechanics and thermodynamics by generating a wavefunction ensemble compatible with quantum statistical mechanics.
Indeed, Kobayashi \cite{Kobayashi96} describes thermal equilibrium of a closed system as the result of a relative-phase interaction.
The relative-phase interaction is simply postulated, however.
While Dugi\'{c} \cite{Dugic97} claims that such an interaction can only happen for open systems, in the dualist interpretation the randomization of relative-phases is a consequence of the P-W guidance condition.
The thermodynamic behaviour of the dualist interpretation has no analogue in the pilot-wave interpretations, whether deterministic or stochastic.

To the second objection, we would respond first by saying that the dualist interpretation resolves the question of the origin of the non-unitary stochastic process of the wavefunction.
In the GRW model, the discrete hits are simply postulated, while in the continuous versions of the EWC interpretation \cite{Pearle89,Diosi.et.al89,Pearle99}, a random field is assumed to exist which interacts with the wavefunction in such a way as to induce a collapse.
In the dualist interpretation, the origin of the non-unitary evolution is attributed to the particle.
In that sense, the dual ontology and the dual guidance condition complement one another.
Furthermore, the GRW model requires that we determine three new physical constants, which must be carefully chosen so that macroscopic objects collapse fast enough (a requirement of its monist ontology) while avoiding excessive energy production.
While the CSL model, on the other hand, demands that we accept the existence of two different types of realities as a means of overcoming the tail problem.
The dualist interpretation, however, is basically a no-collapse interpretation, since objects, microscopic as well as macroscopic, have definite positions at all times.
Consequently, the tail problem simply does not occur, while equilibrium steady states are allowed to exist.
Finally, the concepts of wave and particle are more intuitive than that of `objective' and `projective' realities.

\subsection{Experiments and applications}
There is every reason to think that the experimental results of orthodox quantum theory are reproduced by the dualist interpretation, particularly with respect to microscopic systems comprising a limited number of particles.
This is partly because of the state preparation prior to the experiment, and partly because of the very short duration of the experiment $T$ with respect to frequency parameter, $\lambda^{-1} \approx 10^{16} {\rm s}$.
In an experiment like the one by Arndt et al.\cite{Arndt.et.al99}, large molecules, made up of $N \approx 10^{3}$ particles each, originate from a source in thermal equilibrium, ensuring that they are initially in quantum equilibrium.
The time-of-flight from the source to the detector is of the order $T \approx 10 {\rm ms}$.
The probability that a STE occurs between the source and the detector is $N \lambda T \approx 10^{-15}$.
Therefore, about one large molecule in $10^{15}$ may not exhibit the proper quantum statistics, a statistically insignificant effect.

This is not to say that the dualist interpretation can have no experimental consequences.
The irreversible evolution of systems towards a DQE steady state may provide insight into non-equilibrium thermodynamics, which may also lead to the experimental verification of the dualist interpretation.
The program would consist of: (\emph{i}) developing the dualist interpretation in more detail than what was presented here, (\emph{ii}) finding its consequences (if any) for the non-equilibrium thermodynamics of bulk matter, and finally (\emph{iii}) comparing these results to the observed properties of such systems.
While it is too early to tell if such an experimental program would be feasible, it at least has the advantage of focusing on table-top experiments on bulk matter rather than on the possible detection of the very minute amounts of energy produced by the GRW collapse mechanism \cite{SquiresPearle94}.

\subsection{Speculations and future work}
We must remain open to the possibility that $\lambda$ is not a constant at all, but may vary with particle type, particle mass, or may even be a function of the constants of the wavefunction, $\lambda = \lambda(|C_{0}|,|C_{1}|,...,|C_{K}|)$.
One interesting possibility would be that the frequency parameter is proportional to the energy spread of the wavefunction,
\begin{equation}
\lambda \propto \Delta E /\hbar
\label{E_etu}
\end{equation}
where $\Delta E$ is the standard deviation of the energy of the wavefunction.
Not only would Eq. (\ref{E_etu}) give new meaning to the energy-time uncertainty relation, but it would also imply that energy eigenstates ($\Delta E = 0$) never undergo STEs ($\lambda^{-1} = \infty$).
Such a result would be logical: since STEs have no effect on energy eigenstates, we might as well eliminate them altogether for such states.

The particular version of the dualist interpretation presented here is tentative in many respects.
First and foremost, the stochastic process of discrete transition events should be replaced by a continuous stochastic process on the wavefunction guided by the particle.
This would allow for a greater flexibility than the separable/non-separable criteria used to determine which subset of particles may evolve independently of another subset of particles for the many-particle case.
The phase vector may be subject to a Langevin equation where the drift function would explicitly depend on the particle positions.
The drift function for the phase vector of a given subsystem, therefore, would depend more or less strongly on the particle positions of an external environment depending on the degree of interaction or entanglement of the subsystem with its environment.
Indeed, extending the dualist interpretation to include open systems or systems with a time-dependent Hamiltonian would be as important as developing a continuous stochastic process for the wavefunction.
It also goes without saying that a dualist treatment of quantum fields, relativistic quantum mechanics, spin, to name a few, must be developed.
Above all, a fully comprehensive theory of the wave-particle interaction must be formulated.
In particular, is another conservation principle possible instead of the MSC principle?
Might the constants of the motion be allowed to randomly fluctuate as a result of a STE, but be conserved on average.
In other words, can we replace the strict conservation with a statistical one?
This would certainly be consistent with the stochastic approach developed here.
Alternatively, must all the constants of the motion of the Schr\"{o}dinger equation be conserved?
Perhaps the STEs only conserve those quantities that are constant for the equivalent classical system (total momentum, total energy, and so on).
Such a \emph{Classical Strict Conservation} (CSC) principle would more readily ensure the proper classical trajectories in the macroscopic limit than the MSC principle.
The CSC principle also means that entanglements between energy eigenstates would decay over time, independently of any environmental decoherence.
But due to the very low frequency parameter, that decay should not alter experimental results on microscopic systems.

\section{Conclusions}
\label{S_Concl}

We have presented a version of quantum mechanics in which both the particle and wavefunction are assumed to be objectively real, and where the wavefunction and the particle guide the evolution of the other.
The wavefunction guides the particle according to the stochastic pilot-wave theory.
The particle guides the wavefunction by means of discrete spontaneous transitions that are Poisson distributed in time.
The transitions are assumed to respect the constants of the motion of the Schr\"{o}dinger equation.
Consequently, only the relative phases between the stationary states that make up the non-stationary wavefunction may change as a result of the transitions.
In this way, an action-reaction principle is established between wave and particle.
It is shown that for microscopic objects, the transitions occur so infrequently as to make the dualist interpretation indistinguishable from the orthodox interpretation.
For macroscopic objects, however, the transitions occur so frequently as to cause a rapid decoherence of the wavefunction in the energy basis.
For a free macroscopic body, we have shown that this decoherence causes the emergence of an average classical motion for the center-of-mass of the body.

On a conceptual level, we have argued that the dual ontology and the dual guidance condition complement one another.
The wavefunction and the particle position are now equally important, as they are both necessary to completely specify the stochastic evolution of the wave-particle system.
The dual ontology provides a clearer account of the determinateness of the macroscopic world and quantum measurements than that given by explicit wavefunction collapse models. 
The dual guidance condition creates the genuine decoherence necessary to eliminate empty wave effects and to reproduce classical trajectories in the macroscopic limit.
This kind of decoherence is problematic in pilot-wave theories since decoherence can only be an expression of ignorance of the initial conditions of the wavefunction and is therefore not real.
The dualist interpretation, therefore, imposes a kind of symmetry between the wavefunction and the particle.
Furthermore, since the P-W guidance condition causes a non-stationary pure state to become a steady mixed one, it is argued that the dualist interpretation can exhibit thermodynamic behaviour.
While much work still remains, we nevertheless conclude that the dualist interpretation not only avoids the problems of other interpretations, but can lead to the rigorous unification of classical mechanics, quantum mechanics and thermodynamics.

\end{document}